\documentclass[journal,onecolumn,12pt]{IEEEtran}

\usepackage{graphicx}
\usepackage{cite}   
\usepackage{amsmath,amssymb}
\newtheorem{theorem}{Theorem}
\newtheorem{corollary}{Corollary}
\newtheorem{proposition}{Proposition}
\newtheorem{lemma}{Lemma}
\newcommand{\lcm}{\rm{lcm}}

\begin{document}

\title{A New Construction for LDPC Codes using Permutation Polynomials
  over Integer Rings} 

\author{Oscar Y. Takeshita  \\
Dept. of Electrical and Computer Engineering \\
2015 Neil Avenue \\
The Ohio State University \\
Columbus, OH 43210 \\ 
Takeshita.3@osu.edu\\
\vspace{2em}
Submitted to IEEE Transactions on Information Theory\\
}
\date{\today}

\maketitle

\begin{abstract}
A new construction is proposed for low density parity check (LDPC) codes using
quadratic permutation polynomials over finite integer rings. The
associated graphs for the new codes have both algebraic and
pseudo-random nature, and the new codes are quasi-cyclic. Graph
isomorphisms and automorphisms are identified and used in an efficient search for good
codes. Graphs with girth as large as 12 were found. Upper bounds on
the minimum Hamming distance are found both analytically and
algorithmically. The bounds indicate that the minimum distance grows with block length. 
Near-codewords are one of the causes for error
floors in LDPC codes; the new construction provides a
good framework for studying near-codewords in LDPC codes. Nine example
codes are given, and computer simulation results show the
excellent error performance of these codes. Finally, connections are made 
between this new LDPC construction and turbo codes using interleavers generated by 
quadratic permutation polynomials.    
\end{abstract}


\pagebreak

\section{Introduction}

This paper addresses a new construction for the class of
capacity-approaching low density parity check (LDPC) codes invented by
Gallager~\cite{gal:it62}. However, we start by introducing a result in
turbo coding~\cite{sun:tak:pp}, which motivated this work. Although
the focus is on LDPC codes, we will establish a bridge between turbo
code and LDPC code constructions using quadratic permutation
polynomials (QPP) over finite integer rings. The algebraic structure and excellent
error performance of the new construction are investigated.

Turbo codes using interleavers generated by permutation polynomials
over finite integer rings~\cite{sun:tak:pp} have proven to be both practical
and yield excellent error
performances~\cite{tak:mcf,ryu:tak:qinv,jpl2}. Moreover, the
construction allows a high degree of parallel processing of turbo
decoding without memory access 
contention~\cite{tak:mcf} during the flow of extrinsic information.
To the best of our
knowledge, this is the only known class of algebraic interleavers 
with a very simple description producing turbo codes with error rate
performance meeting or exceeding all other algebraic and pseudo-random 
constructions at practical error rates~\cite{tak:mcf} for a wide range
of block lengths (256--4096-bit interleavers) such as in the
3GPP standard~\cite{3gpp}. We believe that the excellent error rate 
performance is due to two main features of the interleaver: a
``pseudo-randomness'' property obtained by a non-linear
nature~\cite{tak:cos:it} of permutation polynomials of degree two or
larger; and, an algebraic structure allowing designs matched to the constituent
convolutional codes~\cite{sun:tak:pp}. The new construction for LDPC
codes using permutation polynomials in this paper also enjoys this
coexistence of pseudo-randomness and algebraic structure.

LDPC codes, like turbo codes, were initially designed with random
constructions~\cite{mackay:it99,ric:shk:urb:it01,lub:mit:shk:spl:it01,chu:for:ric:urb:cl2001}. 
A ``randomness'' was perceived as an important feature for both types
of codes. However, requirements such as ease of implementation,
large girth for the associated graph, and better error performance
quickly spawned other construction methods. Many good algebraic,
combinatoric, and geometric constructions for LDPC codes have been
proposed in the 
literature~\cite{margulis,ros:von:margulis,kou:lin:fos,fos:geo,sma:von:qcb}. Additionally,
good algorithmic constructions have also been
proposed~\cite{hu:peg}. Most constructions focus on the maximization of the
girth of the associated graph and employ computer simulations for
validation. We follow the same path in this paper.

Recently, some attention has been given to the error floor of LDPC
codes~\cite{richardson:al41,mackay:pos:near}. 
As an example, the $(2640,1320)$ Margulis
code~\cite{ros:von:margulis} has a floor at a frame error rate (FER)
of $10^{-6}$. The two main causes of the floors in LDPC codes are
known to be a small minimum distance and low-weight
near-codewords. Graph symmetries and automorphisms are key properties 
being used to investigate error floors in LDPC
codes~\cite{richardson:al41} below the reach 
of Monte Carlo simulations. For the $(2640,1320)$ Margulis code,
rather than a poor minimum distance\footnote{However, the minimum
  distance of this code is also not impressive because there are codewords of
  Hamming weight 40 for this code~\cite{hu:fos:ele:nncs}.}, the main cause for its floor has been 
identified as low-weight near-codewords of the type $(12,4)$, i.e.,
near-codewords of weight 12 and syndrome weight 4. In our
simulations, example codes for the new construction are given with no apparent
error floors down to FER's close to $10^{-7}$. This does not mean the new codes do not 
have near-codewords; on the contrary, we have identified
near-codewords in one of our example codes. However, their simple
algebraic structure allows an easy identification of graph
automorphisms and may be a valuable framework for the understanding of
error floors in LDPC codes.

The existence of graph automorphisms of our new construction also
implies that permutation polynomial-based LDPC codes are
quasi-cyclic~\cite{fos:itldpcqc}. However, the new construction does not
necessarily generate codes that are equivalent to the codes
in~\cite{fos:itldpcqc}. It is known that the parity check matrices of
quasi-cyclic codes can be written as adjoined circulant square matrices up
to a code equivalence. A recent work~\cite{sma:von:qcb} defines
two subclasses of quasi-cyclic codes. Type I and II codes have equivalent
parity check matrices whose circulant sub-matrices have row/column
weights 0,1 and 0,1,2 respectively. The new construction generates
codes of both types. Codes of
type II were shown in~\cite{sma:von:qcb} to have better minimum
distance upper bounds than codes of type I, which is a superclass of
the codes
in~\cite{fos:itldpcqc}. We further observe a generalization of the result
in~\cite{sma:von:qcb} showing that codes 
generated by the new construction have upper bounds on the
minimum distance potentially growing with the block length. This happens because
the so-called ``weight matrices'' for the circulant sub-matrices of the
parity check matrix have larger sizes (and consequently smaller
circulant sub-matrices) than the previously known constructions. Larger
minimum distances are also confirmed by using the nearest nonzero
codeword search (NNCS)  method, which finds true codewords of low
weight (with a good likelihood of yielding a codeword of lowest
non-zero Hamming weight) in LDPC codes~\cite{hu:fos:ele:nncs}.  

This paper is organized as follows. In section II, we 
define the new LDPC construction and review a result for quadratic
permutation polynomials\cite{sun:tak:pp,ryu:tak:qinv} over the
finite integer ring $\mathbb{Z}_N$. The main results are derived in section
III, and examples and computer simulation results are given in section
IV. Finally, conclusions are discussed in section V.

\section{LDPC Construction}

Let $G$ be a $(\lambda,\rho)$ regular bipartite graph. The graph $G$
consists of $n$ variable nodes $\Lambda=\{v_0,v_1,\ldots, v_{n-1}\}$
with degree $\lambda$ and $r$ check nodes $\Gamma=\{c_0,c_1,\ldots,
c_{r-1}\}$ with degree $\rho$. There are
$N=n\lambda=r\rho$ edges $\Xi=\{e_0,e_1,\ldots,e_{N-1}\}$ in $G$. Each
edge $e_i,0\leq i<N$ has a left-label $i$ and right-label $f(i)$,
where $f(\cdot)$ is a permutation on $\{0,1,2, \ldots, N-1\}$.
Naturally, if the right-label of an edge $e_i$ is $0\leq j<N$ then the
left-label is $i=f^{-1}(j)=g(j)$, i.e., $g(\cdot)$ is the inverse
permutation function of $f(\cdot)$. Each variable node $v_m,0\leq m<n$
is connected to $\lambda$ edges whose left-labels are in the set
$\lambda_m=\{m\lambda, 
m\lambda+1,\ldots, m\lambda+(\lambda-1)\}$. Each check node $c_m,0\leq
m<r$ is connected to $\rho$ edges whose right-labels are in the set
$\rho_m=\{m\rho, 
m\rho+1,\ldots, m\rho+(\rho-1)\}$. Thus every regular $(\lambda,\rho)$
graph $G$ with $N$ edges is completely and uniquely (up to a graph
isomorphism) defined by a permutation $f(\cdot)$ on
$\{0,1,\ldots,N-1\}$. In this paper, we investigate an LDPC
construction when $f(\cdot)$ is a quadratic permutation polynomial
over integer rings~\cite{sun:tak:pp,ryu:tak:qinv}.

In this paper, let the set of primes be $\mathcal{P} = \{p_2
= 2, p_3 = 3, p_5 = 5 , ...  \}$.  Then an integer $N$ can be factored
as $N = \prod\nolimits_{p_i \in \mathcal{P}} p^{n_{N,i}}_i $, where
$p_i$'s are distinct primes, $n_{N,i} \geq 1$ for a finite number of
$i$ and $n_{N,i}=0$ otherwise.
The necessary and sufficient condition for a quadratic polynomial
$f(x)$ to be a permutation polynomial is given in the following
proposition.  

\begin{proposition}\cite{ryu:tak:qinv}\cite{sun:tak:pp}
 Let $N = \prod\nolimits_{p_i \in \mathcal{P}}  p^{n_{N,i}}_{i} $.
The necessary and sufficient condition for a quadratic polynomial $f(x) = f_1 x + f_2 x^2 \pmod{N}$ 
to be a permutation polynomial can be divided into two cases. 

\begin{enumerate}
\item Either $2 \nmid N$ or $4 |N$ (i.e., $n_{N,2}\not = 1$)\\
$\gcd(f_1, N)=1$ and $   f_2 = \prod\nolimits_{p_i \in \mathcal{P}}  p^{n_{{f_2},i}}_{i}, n_{{f_2},i} \geq 1  $, $\forall i$
such that  $n_{N,i} \geq 1$.

\item $2|N$ and $4\nmid N$ (i.e., $n_{N,2}=1$)\\
$f_1+f_2$ is odd, $\gcd(f_1,\frac{N}{2}) = 1$ and $f_2 = \prod\nolimits_{p_i \in \mathcal{P}}  p^{n_{{f_2},i}}_{i}, n_{{f_2},i} \geq 1$, $\forall i$ 
such that $p_i \neq 2$ and $n_{N,i} \geq 1$.
\end{enumerate}
\label{prop:pp2}
\end{proposition}

\section{Design of Good LDPC Codes}

We propose an efficient search for LDPC graphs with large girth by
avoiding inspection of isomorphic graphs. Additionally, we only
check the girth of a graph by computing the local girth starting from
vertices that belong to different equivalence classes under a graph
automorphism.   

\subsection{Isomorphic Graphs}

The following two propositions identify quadratic permutation
polynomials generating LDPC codes with isomorphic graphs.

\begin{proposition}
The graphs generated by $f(x)=f_1x+f_2x^2$ and
$f^\prime(x)=m\rho+f_1x+f_2x^2$, where $m$ is any integer are
isomorphic.
\label{prop:iso0}
\end{proposition}
\begin{proof}
This is readily seen by the definition of the construction and it
simply corresponds to a difference of a constant $m$ modulo $N/\rho$ in the
indices $c_i$ and $c_i^\prime, 0\leq i<r$  of the check nodes of the two graphs.
\end{proof}

\begin{proposition}
The graphs generated by
$f(x)=f_1x+f_2x^2$ and $f^\prime(x)=(f_1+2m\alpha f_2)x+f_2x^2$, where $m$ is
any integer and $\alpha=\lcm(\lambda,\rho)$ are
isomorphic. 
\label{prop:iso}
\end{proposition}

\begin{proof} First we observe that a graph induced by $f(x+m\alpha)$
  is clearly isomorphic to the one induced by $f(x)$ because it just
  corresponds to a different relabeling of the variable
  nodes. Developing $f(x+m\alpha)$ be obtain

\[
 f(x+m\alpha)=f_1(x+m\alpha)+f_2(x+m\alpha)^2=f_1x+f_2x^2+ 2m\alpha f_2x+\alpha m f_1
+\alpha^2 m^2f_2
\]

\[
 f(x+m\alpha)=(f_1+2m\alpha f_2)x+f_2x^2+\alpha m f_1 +\alpha^2 m^2f_2
\]

and the proposition follows from  Proposition~\ref{prop:iso0}.
\end{proof}

Proposition~\ref{prop:iso} implies that in the search for good
QPP's, the range of search for $f_1$
can be set from 1 to $2f_2\alpha$. This means a small $f_2$ may be
advantageous because it reduces the search space for
$f_1$. Proposition~\ref{prop:iso} can also be interpreted as a 
constrained design rule suggested in~\cite{sun:tak:pp} for turbo
codes. Conversely, the search range proposed therein can be interpreted as a
special case of Proposition~\ref{prop:iso} in which $\alpha=1$.  

\subsection{Automorphic Graphs}

The nature of permutation polynomials makes the graph to have automorphisms.
Hence the determination of the girth of the
graph by exhaustive search is simplified by only examining trees
starting from vertices in $G$ that belong to different equivalence
classes under graph automorphisms. We prove next a theorem showing a
graph automorphism with the help of two lemmas.

\begin{lemma}
Let $u=\gcd(2f_2,N)$. Then the set of edge left-labels
\[
\theta_i=\left\{i,i+\frac{N}{u},i+\frac{2N}{u},\ldots,i+\frac{(u-1)N}{u}\right\}
\]
and $|\theta_i|=u$ for all $i$ forms an equivalence class under the
difference of mapped of right-labels.
\label{lem:eqv1}
\end{lemma}
\begin{proof}
This is seen by observing that 
\[
f(x+\gamma)-f(x)\equiv 2f_2x\gamma +f(\gamma) \pmod{N}
\]
and finding the solutions for
\[
2f_2x\equiv 0 \pmod{N}.
\]
\end{proof}

\begin{lemma}
Let $t=\lcm(N/u,\lambda)/\lambda$. Then the set variable nodes
\[
\{v_i,v_{i+t},v_{i+2t},\ldots,v_{i+(N/(t\lambda))-1}\}
\]
for all $i$ forms an equivalence class under the difference of mapped
right-label edges connected to them. 
\label{lem:eqv2}
\end{lemma}
\begin{proof}
This is a direct consequence of Lemma~\ref{lem:eqv1}.
\end{proof}

\begin{theorem}
Let $\beta= mt$ such that $m$ is the smallest
positive integer that makes $\rho|f(mt)$. Then the set of variable
nodes 
\[
\{v_i,v_{i+\beta},v_{i+2\beta},\ldots,v_{i+(n/\beta)-1}\}
\] 
for all $i$ forms an equivalence class under graph automorphisms.   
\label{th:auto}
\end{theorem}
\begin{proof}
The theorem follows from Lemma~\ref{lem:eqv2}.
\end{proof}

\begin{corollary}
Let $\gamma=\frac{\beta\lambda}{\rho}$. Then the set of check 
nodes 
\[
\{c_i,c_{i+\gamma},v_{i+2\gamma},\ldots,c_{i+(r/\gamma)-1}\}
\] 
for all $i$ forms an equivalence class under graph automorphisms.   
\label{co:auto}
\end{corollary}
\begin{proof}
This follows from Theorem~\ref{th:auto} and the natural periodicity of
the variable and check node labels induced by the permutation
polynomial. 
\end{proof}

Theorem~\ref{th:auto} also leads to an intuitive design rule: by
minimizing the number of equivalent classes, the code may look less
uniform and more random. This is achieved by selecting $f_2$ as small
as possible. Following Proposition~\ref{prop:pp2}, we may select $f_2$ to be the
product of every prime factor of $N$ repeated exactly
once, say this number is $f_{2_{\min}}$.\footnote{To simplify the explanation,
  we are assuming only case 
  1) in Proposition~\ref{prop:pp2}, however, the procedure is easily
  generalized for case 2).} Moreover,
increasing $f_2$ with more factors of $N$ also means $f_2$ approaches
$N$, eventually becoming a multiple of $N$ and hence zero modulo
$N$. The QPP then reduces to a linear
permutation polynomial~\cite{tak:cos:linear} losing the ``randomness''
that the non-linear quadratic polynomial provides.  Our search for
good coefficients confirmed this trend where the girth of the
resulting graph increases with $f_2$ but only up to a certain
point. Often, $f_{2_{\min}}$ is the value that maximizes the girth of
the corresponding graph. A similar fact is described
in~\cite{sun:tak:pp} for turbo codes establishing a close tie between
 LDPC and turbo codes based on permutation polynomials. The main
difference between the constructions of turbo codes and LDPC codes
with the permutation polynomials approach is in the constraints:
cycle-lengths of the constituent codes in turbo
codes~\cite{sun:tak:pp} and the degree distribution in LDPC codes.

\subsection{Quasi-Cyclic Representation}

Theorem~\ref{th:auto} naturally implies the new codes are
quasi-cyclic whose {\em shifting
  constraint}~\cite[p. 185]{Lin-Costello-2nd} is 
$\beta$. Quasi-cyclic codes can have their generator and parity check
matrices represented by circulant sub-matrices. Quasi-cyclic
constructions are interesting because encoding can be performed by
shift-registers~\cite[pp. 256--261]{Peterson-Weldon}. 
Quasi-cyclic LDPC codes from circulant permutation matrices
have some important limiting factors; the girth of the
graph is at most 12~\cite{fos:itldpcqc} and the minimum Hamming
distance of the LDPC code is 
upper bounded by $(\lambda+1)!$~\cite{MacKayHighRate98}. These ideas
have been recently generalized~\cite{sma:von:qcb} when circulant
sub-matrices are allowed to have row/column weights 0 and 1 (type I
constraint) and row/column weights 0, 1 and 2 (type II
constraint). Our construction generates both type I and II codes but
are not equivalent to the construction 
in~\cite{fos:itldpcqc}, which are of type I with circulant matrices of
weight 1 only. The non-equivalence can be demonstrated by a counter example. The
codes in~\cite{fos:itldpcqc} always generate rank deficient parity
check matrices whereas we have observed that in general our
construction yields full rank parity check matrices (examples are
given in Section~\ref{sec:examples}). We show next an
example code that has constraints of type II.  

\begin{figure}[htbp]
\centering
\includegraphics[height=5cm,width=10cm]{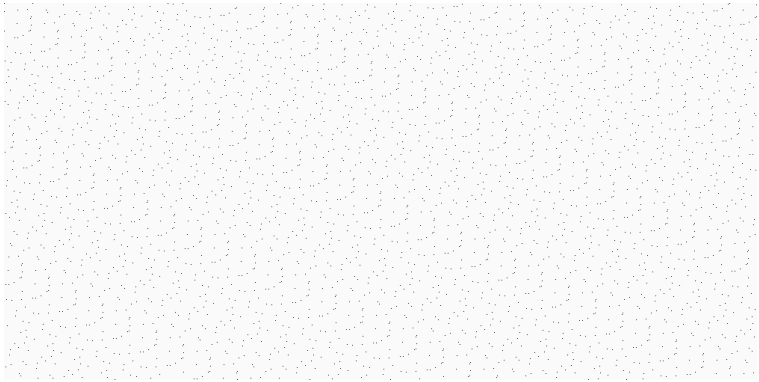}
\caption{Parity check matrix for the (3,6)-regular $(1008,504)$ example
  code II.} 
\label{fig:parityH}
\end{figure}

The example code II in Table~\ref{tab:codes} is (3,6)-regular with
size $(1008,504)$. The plot of its parity check matrix is shown in 
Figure~\ref{fig:parityH}. A dark dot represents a 1 and 0 otherwise. 

The matrix apparently lacks regularity. This is the ``randomness''
introduced by the quadratic permutation polynomial. However,
 the parity check matrix can be rearranged by grouping columns and rows according to their
corresponding variable node and check node equivalence classes using
Theorem~\ref{th:auto} and Corollary~\ref{co:auto}. Let the parity
check matrix be  $H=[x_0x_1\ldots x_{n-1}]$ where $x_i$'s are column
vectors. Then define 

\[
H^\prime=[
x_0 x_{0+\beta}\cdots x_{0+(n/\beta-1)\beta},
x_1 x_{1+\beta}\cdots x_{1+(n/\beta-1)\beta}
\cdots
x_{\beta-1}x_{\beta-1+\beta}\cdots x_{\beta-1+(n/\beta-1)\beta}]
\]

Let now $H^\prime=[y_0,y_1,\cdots,y_{r-1}]^T$, where the $y_i$'s are
column vectors and $T$ denotes matrix transposition. Then define

\[
H^{\prime\prime}=[
y_0 y_{0+\gamma}\cdots y_{0+(n/\gamma-1)\gamma},
y_1 y_{1+\gamma}\cdots y_{1+(n/\gamma-1)\gamma}
\cdots
y_{\gamma-1}y_{\gamma-1+\gamma}\cdots y_{\gamma-1+(n/\gamma-1)\gamma}]^T
\]

\begin{figure}[htbp]
\centering
\includegraphics[height=5cm,width=10cm]{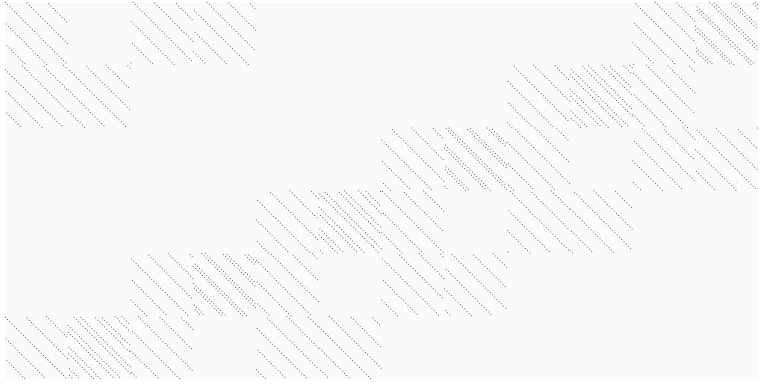}
\caption{Parity check matrix for the (3,6)-regular $(1008,504)$ example
  code II in circulant form.} 
\label{fig:parityHcirc}
\end{figure}

These simple transformations of the parity check matrix result
in a new matrix consisting of circulant sub-matrices of size
$r/\gamma\times n/\beta$. In our example, the circulant sub-matrices
are of size $84\times 84$ and the resulting parity check matrix is
shown in Fig.~\ref{fig:parityHcirc}.   

This means that the new code is quite structured after
all. More precisely, it can be said that the ``pseudo-randomness''
introduced by the non-linearity of the second degree permutation
polynomial has been factored. We will see next where the
``pseudo-randomness''  gets condensed. The weight of the rows of the  
circulant sub-matrices in this example are 0, 1 and 2. This is
depicted by a weight matrix $A$ in (\ref{eq:circweight}).

\begin{equation}
A=\left[
\begin{array}{cccccccccccc}
1 & 0 & 1 & 1 & 0 & 0 & 0 & 0 & 0 & 0 & 1 & 2  \\ 
1 & 1 & 0 & 0 & 0 & 0 & 0 & 0 & 1 & 2 & 1 & 0  \\
0 & 0 & 0 & 0 & 0 & 0 & 1 & 2 & 1 & 0 & 1 & 1  \\
0 & 0 & 0 & 0 & 1 & 2 & 1 & 0 & 1 & 1 & 0 & 0  \\
0 & 0 & 1 & 2 & 1 & 0 & 1 & 1 & 0 & 0 & 0 & 0  \\
1 & 2 & 1 & 0 & 1 & 1 & 0 & 0 & 0 & 0 & 0 & 0  \\
\end{array}
\right]
\label{eq:circweight}
\end{equation}

Observing matrix $A$, the ``pseudo-randomness'' is condensed in the
arrangement of the circulant sub-matrices of $A$. Further, matrix $A$ has
a more general form than the matrices obtained 
in~\cite{MacKayHighRate98,sma:von:qcb}. Both work demonstrate an upper
bound on the minimum distance assuming that the number of rows in $A$
is $\lambda$. The recent work by Smarandache and
Vontobel~\cite{sma:von:qcb} shows that quasi-cyclic constructions of
type II have larger minimum distance upper bounds than the 
ones for type I (a similar result for high-rate quasi-cyclic LDPC
codes is reported in~\cite{kam:hrqc}). The upper bound~\cite[Theorem 2]{sma:von:qcb} and an
example code are provided for $\lambda=3$ achieving the upper bound of
$d_{\min}\leq 32$. Their
theorem is stated as follows: 

\begin{theorem}[\cite{sma:von:qcb}] Let $C$ be a quasi-cyclic code
  with a $\lambda\times \rho$ weight matrix $A$. 
\[
d_{\min}\leq \min_{\substack{S\subseteq\{1,\ldots,\rho\}\\|S|=\lambda+1}} \sum_{\substack{
    S^\prime\subset S\\
    S^\prime=\{i_1,\ldots,i_{\lambda}\}
}}\sum_{\sigma \in \Pi}
a_{\sigma(1),i_1}\cdots a_{\sigma(s\lambda),i_{\lambda}}
\label{th:dminupper}
\]
where $\Pi$ is the set of all permutations of $\{1,\ldots, \lambda\}$
and $a_{x,y}\quad x=1,2,\dots,\lambda \quad y=1,2,\dots,\rho$ are the
entries of $A$. 
\end{theorem} 

MacKay's Theorem 2 in~\cite{MacKayHighRate98} can be interpreted as a
special case of the previous theorem when all entries in $A$ are equal
to 1.  The important point in Theorem~\ref{th:dminupper} 
is that by allowing different weights for the circulant matrices, a larger
upper bound is obtained. We make another
straightforward generalization, which improves the upper bound on the
minimum distance even when the codes are of type I. By allowing $A$ to
have a number of rows larger than $\lambda$ (and columns larger than $\rho$) to accommodate our weight
matrices, we obtain the following theorem.  

\begin{theorem} Let $C$ be a quasi-cyclic code
  with a $s\lambda\times s\rho$ weight matrix $A$. Let $S\subseteq
  \{1,\ldots, s\lambda+1\}$ and 
\[
\psi(S)=\sum_{\substack{
    S^\prime\subset S\\
    S^\prime=\{i_1,\ldots, i_{s\lambda}\}}}\sum_{\sigma \in \Pi}
a_{\sigma(1),i_1}\cdots a_{\sigma(s\lambda),i_{s\lambda}}
\]
where $\Pi$ is the set of all permutations of $\{1,\ldots,s\lambda\}$.
Then 
\[
d_{\min}\leq \min_{\substack{S\subseteq\{1,\ldots,s\rho\}\\
|S|=s\lambda+1\\
\psi(S)\neq 0}}\psi(S)
\]
\label{th:dminupper2}
\end{theorem}
\begin{proof}
This follows directly from Theorem 2 in~\cite{MacKayHighRate98} and Theorem 2
in~\cite{sma:von:qcb}. However, it is important to exclude $\psi(S)=0$,
otherwise the upper bound easily incorrectly evaluates to zero for the
new construction. This is because in the new construction, the weight
matrix $A$ is itself of low density as opposed to the ones
in~\cite{MacKayHighRate98,sma:von:qcb} which are
dense\footnote{Theorem 2 in~\cite{sma:von:qcb} must
  also exclude the instances of $\psi(S)=0$ to be strictly
  correct.}. The event $\psi(S)=0$ happens when the set $S=\{j_1,j_2,\ldots,
j_{s\lambda+1}\}$ defines a 
sub-matrix $A(S)=[b_{j_1} b_{j_2} \ldots b_{j_{s\lambda+1}}]$ containing
a row of zeros where $b_{j_i}$ is the $j_i$th column of $A$.

\end{proof}

Applying Theorem~\ref{th:dminupper2} on matrix $A$, we obtain $d_{\min}\leq
62$ for the new $(1008,504)$ code II, which is greater than the upper
bound of 24 for all type I codes with $\lambda=3$. It is also greater than the
upper bound of 32 for type II codes with a dense weight matrix
in~\cite{sma:von:qcb}. 

 Tighter bounds may be obtained by recursively applying
 Theorem~\ref{th:dminupper2} on the sub-matrices corresponding to the
 events $\psi(S)=0$ but with the all-zeros row removed. As an example,
 in the previous matrix $A$, if $S=\{1,2,3,4,5,6\}$ then $\psi(S)=0$
 because the  corresponding matrix has the third row from the top
 all-zero. We can  thus apply Theorem~\ref{th:dminupper2} on the
 matrix  

 \begin{equation}
 A^\prime=\left[
 \begin{array}{cccccc}
 1 & 0 & 1 & 1 & 0 & 0 \\
 1 & 1 & 0 & 0 & 0 & 0 \\
 0 & 0 & 0 & 0 & 1 & 2 \\
 0 & 0 & 1 & 2 & 1 & 0 \\
 1 & 2 & 1 & 0 & 1 & 1 \\
 \end{array}
 \right]
 \end{equation}

However, the bound was not improved for this code.

Another upper bound on the minimum distance for
code II was computed by using the nearest nonzero codeword search
(NNCS) method in~\cite{hu:fos:ele:nncs}, which is likely to yield
the nonzero codeword of lowest Hamming weight. Codewords of weight 44 have been
found. Therefore Theorem~\ref{th:dminupper2} gives a loose upper bound for this
code. Moreover, no codewords of low weight have been found by extensive
computer simulations in Section~\ref{sec:examples} via undetected
errors. In summary, the combination of circulant matrices of weights 0, 1,
and 2 together with a low density weight matrix gives good minimum
distances for the new code construction. Another interesting example is
code III. Its weight matrix is of type I, therefore the new
construction generates both type I and II codes. Its weight matrix is
of size $64\times 128$, therefore the explicit computation of the
minimum distance upper bound in Theorem~\ref{th:dminupper2} becomes
complex, at least by a brute force.  However, the NNCS method is still
manageable giving an upper bound on the minimum distance of
78. We conjecture the minimum distance
of the new construction grows linearly with the block length. This
fact is supported by the upper bounds on the minimum distance computed
for codes of several lengths as shown in Table~\ref{tab:codes}. We
also observe that the tightness of the upper bound in
Theorem~\ref{th:dminupper2} with the NNCS method gets looser with the
block length. Finally, even the NNCS method has its limitations in
terms of complexity. A full-run of the algorithm using a ``two
position bit reverse'' is very costly for code V. The upper bound of
248 is only for a limited search with some optimization by exploiting
the automorphism of the code.

\subsection{A Search Procedure for Good Coefficients}

We give an outline of an efficient search procedure for graphs
with large girth using Propositions~\ref{prop:iso} and Theorem~\ref{th:auto}.

\begin{enumerate}
\item Choose the degree distribution $(\lambda,\rho)$
\item Choose the code size $(n,k)$ and set the interleaver length $N=n\lambda$
\item Set $f_2$ to be the product of every prime factor of $N$ repeated
  exactly once
\item Search the $f_1$ that maximizes the girth of the corresponding
  graph using Propositions~\ref{prop:iso} and Theorem~\ref{th:auto} 
\item Try a larger $f_2$ by including more factors\footnote{The
  inclusion of factors not present in $N$ is in some cases
  effective. This was the case for code I in Table~\ref{tab:codes}.}
of $N$ and repeat
  the previous step. If no improvement in girth is noted, proceed to
  the next step otherwise 
  repeat this step
\item Compute the rank of the parity check matrix to find the true
  dimension of the code
\end{enumerate}

The search procedure quickly generates on a regular personal computer
all codes in Table~\ref{tab:codes}.

\section{Example Codes and Simulation Results}
\label{sec:examples}

\begin{figure}
  \centering
  \includegraphics[width=0.7\hsize,angle=270]{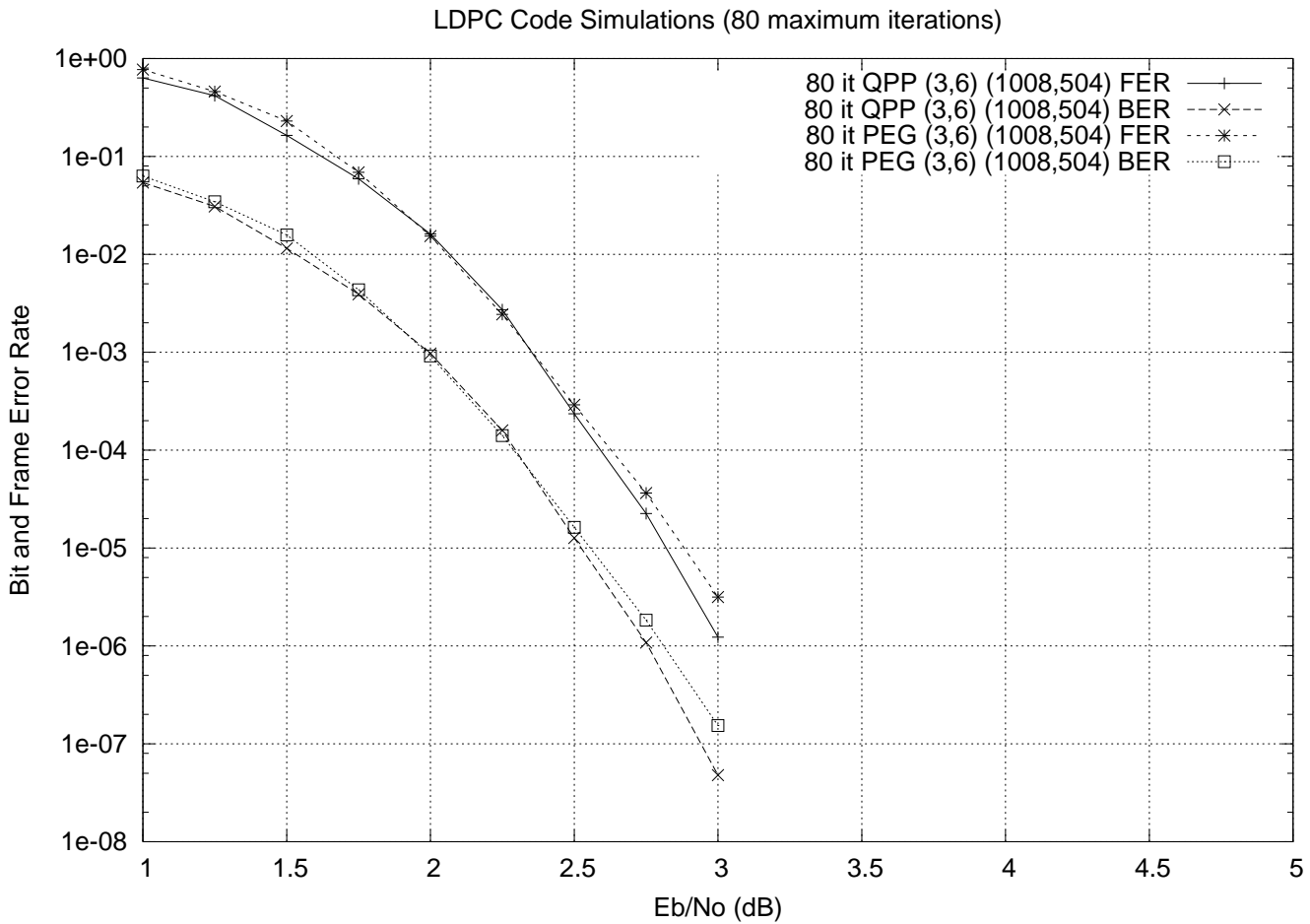}
  \caption{BER and FER curves for the new (3,6)-Regular $(1008,504)$ and PEG
    (3,6)-Regular $(1008,504)$ codes using 80 BP decoding iterations}
  \label{fig:sim1}
\end{figure}

Nine new QPP-based LDPC codes are listed in
Table~\ref{tab:codes}. Most of the entries are self-explanatory. The 
columns $d_{\min}$ and $d_{\min}^{T3}$ represent upper bounds on the
minimum distance computed using the NNCS method in~\cite{hu:fos:ele:nncs} and
Theorem~\ref{th:dminupper2}, respectively. The only code that has rate not equal to
exactly 1/2 is code VIII  because its parity check matrix is
rank deficient. We simulated codes II, IV and VI using BPSK modulation
under an additive white Gaussian noise (AWGN) channel. We first simulated
code II, which is (3,6)-regular and has size $(1008,504)$. It was
compared with a girth-8 progressive edge growth (PEG) code~\cite{hu:peg}. For a fair
comparison with the curve in~\cite{hu:peg}, we used the same number of
80 belief-propagation (BP) decoding iterations. At least 50 frame errors were
counted per simulated point in all of our simulations unless otherwise
noted. Simulation curves for bit error rate (BER) and frame error rate
(FER) are shown in Figure~\ref{fig:sim1}. The new QPP code outperforms
the PEG code at high signal-to-noise ratios (SNR's).   

\begin{table}[htbp]
\centering
\caption{New QPP LDPC codes}
\label{tab:codes}
\[
\begin{array}{|c|c|c|c|c|c|c|c|c|c|} \hline
 \mbox{Code} & (\lambda,\rho) & (n,k) & f(x) &  N & \mbox{girth} &
 \beta & d_{\min}^{} &  d_{\min}^{T\ref{th:dminupper2}} 
\\ \hline
I & (3,6) & (504,252) & 5x+210x^2 &   1512 & 8 &  6  & 22 & 22 \\
II & (3,6) & (1008,504) & 29x+42x^2 &   3024 & 8  & 12 & 44 & 62 \\
III &  (3,6) & (2048,1024) & 7x+24x^2 &  6144 & 8 & 128 & 78 & -\\
IV &  (3,6) & (2432,1216) & 11x+114x^2 &  7296 & 10  & 32 & 90 & 344 \\
V &  (3,6) & (4096,2048) & 43x+24x^2 &  12288 & 10 & 256 & 248 & -\\
VI &  (3,6) & (8192,4096) & 19x+24x^2 &  24576 & 10  & 512 & - & - \\
VII &  (3,6) & (16384,8192) & 7x+24x^2 &  49152 & 10  & 1024 & - & - \\
VIII &  (3,6) & (32768,16384^*) & 7x+48x^2 &  98304 & 12 & 1024 & - & -\\
IX &  (4,8) & (1120,562)  & 87x+70x^2 &  4480 & 8 &  8 & 40 &  96\\  \hline
\end{array}
\]
* The true dimension has not been computed.
\end{table}

\begin{figure}
  \centering
  \includegraphics[width=0.7\hsize,angle=270]{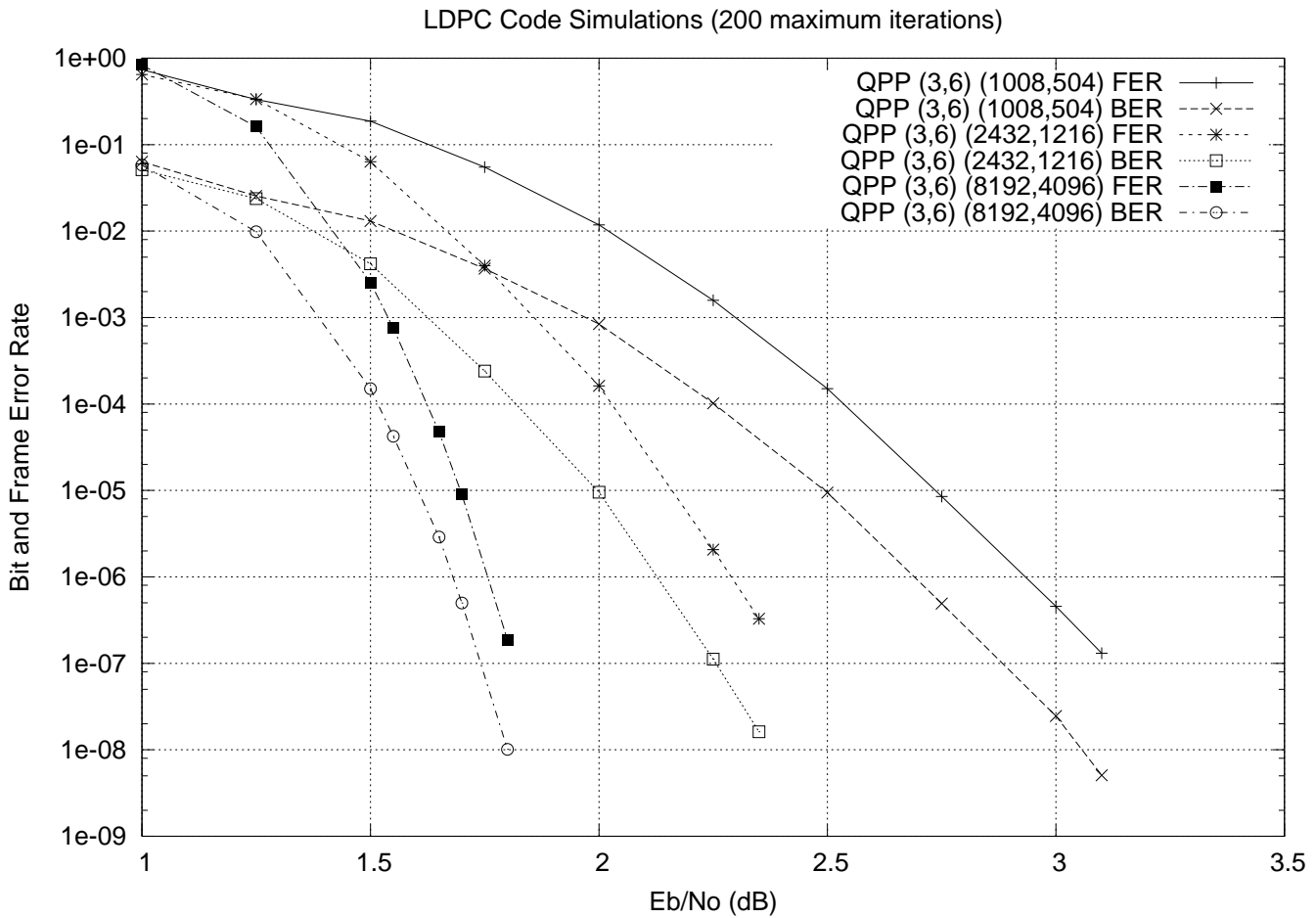}
  \caption{BER and FER curves for the QPP LDPC codes II, IV and  VI using 200
    BP iterations} 
  \label{fig:sim}
\end{figure}

We next extensively simulated codes II, IV and VI using 200
BP decoding iterations. The BER and FER simulation curves are
shown in Figure~\ref{fig:sim}.

\subsection{Code II (3,6)-Regular $(1008,504)$}

For code II, we simulated over 168 million
frames at 3.00dB and accounted for 77 detectable frame errors but no
undetectable frame errors. We also found that among the detected frame
errors there were 8 (10\%) low-weight near-codewords of the type $(12,2)$, i.e.,
a near-codeword of weight 12 and syndrome of weight 2, respectively. 
Over 146 million frames were simulated at 3.10dB. Out of the 19
detectable frame errors, 8 (42\%) were $(12,2)$
near-codewords. This is a significant
increase in the percentage of low-weight near-codewords out of the
detected frame errors compared with the results at the SNR of
3.00dB. This indicates an error floor to appear at an FER near
$5\times 10^{-8}$. Therefore, code II has a similar problem
with near-codewords as the Margulis $(2640,1320)$ code~\cite{margulis} 
as reported in~\cite{mackay:pos:near}. However, the floor is about one order of
magnitude below that of the Margulis code.

\subsection{Code IV (3,6)-Regular (2432,1216)}

Because code II has only half the length of the Margulis $(2640,1320)$ code
studied in~\cite{mackay:pos:near}, code IV was simulated for a fair
comparison. The results are very encouraging because there are no 
signs of an error floor down to an FER of $3\times
10^{-7}$. Further, the following indicates that a floor must be much
lower. Over 16 million frames were simulated at 2.25dB resulting in a
total of 34 detected frame errors. Near-codeword and
syndrome weights were mostly triple-digit except for 6 of them with
the following weights: $(16,40)$, $(47,95)$, $(80,156)$, $(83,99)$, $(84,140)$,
and $(93,125)$. Over 63 million frames were simulated at 2.35dB
resulting in a total of 21 detected frame errors. Near-codeword and
syndrome weights were mostly triple-digit except for 3 of them with
the following weights: $(45,99)$, $(84,118)$, and $(96,128)$.  
 
\subsection{Code VI (3,6)-Regular (8192,4096)}

Code VI has a block size of approximately 8000 
bits as all the codes compared at the Jet Propulsion Laboratory (JPL)
in~\cite{and:dol:div:tho:jpl42-159}. There are two (3,6)-regular codes
compared in~\cite{and:dol:div:tho:jpl42-159}: a random code and a PEG
code. Both codes do not have structure and easy encoding and decoding
methods. The PEG code outperforms the random 
code but the new code IV performs no worse than the PEG
code. Further, encoding and decoding of the new code are simpler due to its
algebraic structure. We observed no low-weight near-codewords at
1.70dB for over 3 million simulated frames and 28 detected frame
errors (FER = $8.9\times 10^{-6}$). The number of frames simulated at
1.80dB was over 16 million with 3 detected frame errors (FER =
$1.7\times 10^{-7}$). All detected frame errors at 1.70dB and 1.80dB
had triple-digit near-codeword and syndrome weights.

\section{Conclusions}

We proposed a new construction for regular LDPC codes using quadratic
permutation polynomials (QPP) over finite integer rings. It is one of
the simplest  
known constructions and yet provides enough flexibility to generate
a large family of good codes of practical interest. The new
construction only requires the code size $(n,k)$, the degree
distribution $(\lambda,\rho)$ of the associated bipartite graph with
$N$ edges, and two integers representing the coefficients of a
QPP $f(x)=f_1x+f_2x^2\pmod{N}$ as the
code defining parameters. The algebraic structure allows the
identification of graph isomorphisms and automorphisms, which
significantly simplify the search for good parameters for the 
coefficients of $f(x)$. The degree of $f(x)$ being two ensures a
non-linearity in the graph structure, which we believe makes it as
good as random constructions with the additional advantages of an
algebraic construction. LDPC codes with corresponding bipartite
graphs with girth as large as 12 are easily obtained. Analytical and
algorithmic upper bounds on the minimum distance were given and computed
for the new codes. Although not formally proven, the new codes appear
to have a minimum distance growing with the block length as opposed to
a fixed small upper bound on the minimum distance of 24 for
quasi-cyclic LDPC codes in~\cite{fos:itldpcqc}. Simulation results
confirm that the new codes have excellent error rate 
performance. In particular, we have found neither undetected errors
nor noticeable error floors down to frame error rates close to
$10^{-7}$; therefore, the new QPP LDPC codes exceed the performance of other
algebraic constructions such as the Margulis
construction~\cite{margulis,ros:von:margulis}. We do expect an error
floor around an FER of $10^{-8}$ for 
one of our codes caused by low-weight near-codewords. However, because the new
codes have easily identifiable graph automorphisms, we believe a
framework can be set for a further understanding of the important
issue of error floors in LDPC codes due to low-weight near-codewords using the
techniques in~\cite{richardson:al41,mackay:pos:near}.

We also narrowed the theoretical gap between LDPC codes and
turbo codes under the unified method of permutation polynomial
algebraic interleavers over integer rings. The designs of the
permutation polynomials were very similar. 
The constraints for an LDPC code were the degree distribution,
while for a turbo code the constraints were the cycle-length of the recursive
constituent codes.

In this paper, only regular constructions have been
demonstrated. However, an extension to irregular codes is possible by
laying out vertex labels periodically according to their corresponding
node degrees. This will be investigated in future work. 

Finally, from the practical side, the new QPP LDPC codes have very attractive
features. They can be encoded by shift-registers because they are
quasi-cyclic, and a parity check matrix in the form of circulant
sub-matrices is easily obtained. Moreover, the decoding allows a high
degree of parallel processing without exhibiting memory access
contention~\cite{tak:mcf} caused by extrinsic information flow.     

\section{Acknowledgement}

The author wishes to thank Marc Fossorier for stimulating
discussions. The author also thanks David MacKay for discussions and
for providing additional data clarifying the low-weight
near-codewords identified in~\cite{mackay:pos:near} for a Margulis
code.  

\bibliography{myresearch}
\bibliographystyle{IEEEbib}
\end{document}